# Fast non-iterative algorithm for 3D point-cloud holography

NATHAN TESSEMA ERSARO,[1,†,*] CEM YALCIN,[2,†] LIZ MURRAY,[2] LEYLA KABULI,[2] LAURA WALLER,[1,2] AND RIKKY MULLER[1,2]

[1]*Graduate Program in Bioengineering, University of California, Berkeley and University of California, San Francisco, Berkeley, CA 94720, USA*
[2] *Department of Electrical Engineering and Computer Sciences, University of California, Berkeley, Berkeley, CA 94720, USA*
[†]*These authors contributed equally to this work*
*\*nathan_ersumo@berkeley.edu*

**Abstract:** Recently developed iterative and deep learning-based approaches to computer-generated holography (CGH) have been shown to achieve high-quality photorealistic 3D images with spatial light modulators. However, such approaches remain overly cumbersome for patterning sparse collections of target points across a photoresponsive volume in applications including biological microscopy and material processing. Specifically, in addition to requiring heavy computation that cannot accommodate real-time operation in mobile or hardware-light settings, existing sampling-dependent 3D CGH methods preclude the ability to place target points with arbitrary precision, limiting accessible depths to a handful of planes. Accordingly, we present a non-iterative point cloud holography algorithm that employs fast deterministic calculations in order to efficiently allocate patches of SLM pixels to different target points in the 3D volume and spread the patterning of all points across multiple time frames. Compared to a matched-performance implementation of the iterative Gerchberg-Saxton algorithm, our algorithm's relative computation speed advantage was found to increase with SLM pixel count, reaching >100,000x at 512x512 array format.

## 1. Introduction

The dynamic patterning of 3D optical point clouds – sparse collections of target points within a volume - has emerged as a key enabling technology in photosensitive volumetric processing across several applications. In biological microscopy, 3D point cloud patterning is employed for non-invasive all-optical interfacing with cell ensembles such as neuronal populations [1–4]. Such interfacing is enabled by bioengineered cell membrane read/write probes, namely fluorescent genetically encoded voltage indicators (GEVIs) that track individual firing events [5] and opsin proteins that respond to light by modulating neuron firing [3]. In material processing, point cloud patterning can also be mobilized for 3D fabrication via lithographic photopolymerization [6–8]. Several shared commonalities and considerations can be identified in such applications involving optical point cloud patterning across a photoresponsive medium. First, achieving diffraction-limited resolution is often not critical as target spot confinement is aided by auxiliary techniques, including multiphoton excitation [6,9,10] and temporal focusing [3,11,12]. In biology, target illumination spots may also be expanded to span whole cells in order to simultaneously excite all membrane-bound probes across a given target cell [13,14]. Second, most 3D intensity distributions cannot be perfectly reproduced due to optical propagation and energy conservation principles [15,16]. In the case of 3D point clouds, beamlets targeting distinct target points introduce undesirable illumination to non-target regions across the patterned volume as they converge to and diverge from target points. Therefore, increasing the density of target points in a given patterned volume reduces contrast, defined as the ratio of irradiance in target regions to irradiance in non-target regions, thus raising the likelihood of undesirable polymerization photoreactions in 3D nanofabrication [17–

20] or the activation of off-target cells in biological microscopy [3,9]. Third, high-power requirements for multiphoton excitation and temporal focusing approaches create a power-limited regime of operation, bottlenecking the total number of points that may be simultaneously addressed [6,9]. These contrast and power limitations typically impose a sparsity constraint that limits the total number of points simultaneously addressed in the target volume.

Ideally, serial 3D single-point scanning approaches can serve to circumvent power limitations and maximize contrast if there are no bottlenecks to optical patterning and photoresponse speeds [6,14]. However, higher optical target power can often not be traded away for near-instantaneous illumination time due to constraints involving photoresponse kinetics, photobleaching, and target heating [21], prohibitively constraining targeting throughput, defined as the number of distinct target points addressed within a given time frame. For instance, opsins and GEVIs require millisecond-scale dwell times that limit patterning rates to kilohertz regimes [22,23]. Therefore, parallel illumination, involving the simultaneous addressing of multiple points with each patterning frame, is often critical to meaningful targeting throughputs that allow for the interrogation of complex biological tissue dynamics or faster fabrication cycles. Typically, such parallel 3D optical point cloud patterning is achieved with computer-generated holography (CGH) powered by spatial light modulators (SLMs). These dynamically-configurable arrayed surfaces impart pixel-level modulation to an incident laser beam in Fourier space in order to generate a desired 3D intensity distribution in image space [3,24]. Specifically, SLMs that offer pseudo-analog phase control are better suited for efficient optical patterning as they do not inherently impart loss through their modulation mechanism [25,26]. Phase also captures more of the information for intensity distributions, especially for high-frequency intensity patterns such as sparse point clouds, in the Fourier domain [15,27].

Critically, achieving closed-loop operation with such optical point cloud patterning to realize systems involving read/write brain-machine interfaces [2,4] or camera monitoring and feedback for photopolymerization-based fabrication [10,28] requires real-time CGH computation. Look-up-table-based methods construct holograms by superimposing the fringe phase patterns that are associated with each target location from stored memory [29,30]. Since such approaches require exorbitant storage and data transfer speeds, strategies to reduce memory usage via compression, quality reduction of the 3D representation map, or additional real-time computation have also been developed [18,29–31]. However, these approaches remain nonviable for real-time deployment as overall computational burden remains high when time complexity is traded off against space complexity: a full SLM-size 2D matrix must fundamentally still be generated for each point [18,31]. Furthermore, the discrete 3D representation maps of accessible point locations preclude the possibility of arbitrary positioning within the target volume, thereby limiting patterning quality [18]. Alternatively, efforts to alleviate computation for each point location via a wavefront recording plate located near enough to the target volume that propagations from separate targets do not overlap mandate a small depth range, severely constraining the accessible volume for patterning [E, F]. Additionally, while single-point scanning remains easy and fast to compute under CGH using deterministic focus tuning and lateral steering relationships [31,32], superimposing single-point SLM masks together to target multiple points with a single SLM frame fails to perform well under phase-only control for high-frequency intensity patterns such as sparse point clouds [33].

Thus, CGH computation warrants iterative approaches, of which the most straightforward and widely used implementation is the Gerchberg-Saxton (GS) algorithm. GS computationally propagates complex fields back and forth between SLM and target volume planes while enforcing amplitude constraints to converge on a hologram for a given point cloud distribution [16,34]. Other gradient descent-based iterative approaches may also be employed for improved hologram optimization via custom penalties at the cost of more computation [35,36]. However, iterative methods remain prohibitive to real-time computation as each propagation step requires a computationally expensive, full SLM-size fast Fourier transform (FFT) operation. For an SLM of pixel count $F \times F$ targeting a volume discretized into $M$ planes, each iteration's resulting computational time complexity is $O(MF^2 \log(F))$. Deep learning approaches using convolutional neural networks (CNNs) offer a promising solution to fast non-iterative CGH computation, with recent efforts achieving computations times on the order 10-100 ms [37,38]. However, CNNs require hardware-intensive graphics processing units (GPUs) or accelerators, preventing real-time deployment in hardware-light settings such as mobile systems where computation times can reach seconds [38]. Furthermore, similar to iterative algorithms, the computation times of CNN-based CGH algorithms are heavily impacted by the resolution of the volumetric representations employed in computation. Axially, existing CNN-based approaches drastically limit addressable depths to ~10 distinct planes, precluding the possibility of truly arbitrary target point placement across the allowable 3D field of view (FoV) [35,36]. Since lateral resolution is given by SLM pixel count, CNN-based CGH algorithms may also fail to retain their speed advantage with future generations of ultra-high-resolution SLMs without commensurate hardware improvements [39].

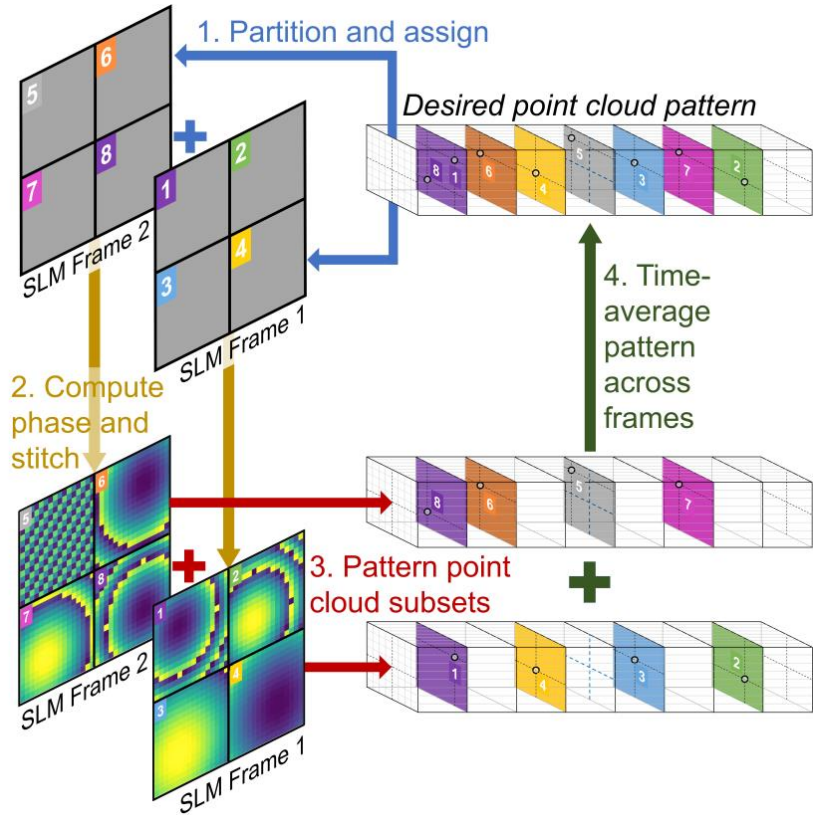

Fig. 1. Diagram illustrating the NIMBLE-PATCH algorithm's non-iterative, patchwork approach to phase mask computation. The SLM array is first partitioned into multiple evenly-sized patches across a number of time frames based on field of view, spot size, and targeting throughput specifications. For a given target point cloud pattern, each target position is assigned to a separate SLM patch such that overall diffraction efficiency is maximized. Finally, phase masks are computed separately across each patch using deterministic beamsteering and focus tuning relationships, then stitched together to generate full-frame phase masks that can produce the desired point cloud pattern.

Fundamentally, current CGH approaches are not ideally suited to sparse 3D point cloud patterning due to an inefficient allocation of available degrees of freedom to target points in the CGH computation process. For a 3D point cloud consisting of $T$ target points, each point located at a unique depth separately adds expensive $F \times F$ matrix-wide computation spanning the full SLM. Yet each target point is allocated only a fraction $1/T$ of both the total available optical power and the SLM étendue available for modulation, which is proportional to the total SLM pixel count $F \times F$ [33,40]. Since sparse 3D point clouds are formed by a finite set of distinct beamlets (one for each target point) whose mutual interference results in minimal impact on image quality, judicious partitioning of the full set of SLM pixels into distinct groups allocated to each target point can serve to alleviate computational burden [41]. The beamlets originating from each group of SLM pixels can thus be treated as being mutually incoherent, though speckle artifacts from mutual interference may also be mitigated via random phase offsets applied across SLM pixel groups if the SLM speed can accommodate time-averaged operation [42–45]. Interleaving single-point holograms across the full SLM frame offers the most straightforward path to such SLM partitioning and allocation [46]. However, interleaved phase masks preserve diffraction-limited resolution at the expense of native field of view, which is already unduly constrained by the large pixel pitches of existing SLM offerings relative to the operating wavelength [9,25,26]. Therefore, an outstanding need remains for an algorithm that balances the throughput benefits of parallel illumination and the fast computation times of single-point patterning while preserving the ability to place target points with infinitesimal precision across the maximum FoV allowed by the SLM.

Accordingly, we present NIMBLE-PATCH, an algorithm with Non-Iterative, Multi-Block, Local Efficiency-driven Point Assignment and Targeting for Cloud-based Holography. The algorithm's procedure, illustrated in Figure 1, involves initially partitioning the SLM array into several evenly-sized patches (i.e., pixel blocks) per FoV, resolution, and throughput specifications. The full 3D point cloud is subsequently decomposed into subsets of point clouds to be addressed by a number of frames set in accordance with time multiplexing capabilities, with every target point being assigned to a specific SLM patch on its respective frame. To maximize the diffraction efficiency and resulting contrast of the patchwork holograms, the allocation of patches to targets is accomplished by constructing a diffraction loss matrix between each target and each patch center, then solving the associated linear sum assignment problem. The algorithm's FFT-free, efficiency-driven approach thus employs lightweight deterministic phase calculations that scale primarily with target count $T$ for the optimal allocation of available degrees of freedom across targets.

To rigorously evaluate our algorithm's performance, we developed a computational simulation framework that is agnostic to optical system parameters and accounts for volume and resolution scaling across SLM pixel count $F \times F$, target count $T$, and time-multiplexed frame count $N$. This framework was employed to systematically compare NIMBLE-PATCH against GS-based algorithms across $F$ and $T$ to identify performance and computation time trends. Compared against the least computationally burdensome implementation of GS involving minimal sampling of the SLM and target planes, NIMBLE-PATCH reaches double the contrast values of GS at pixel counts as low as 512x512 within compute times that are $>10^4\times$ faster. In addition, an improved implementation of GS matching the contrast performance of NIMBLE-PATCH via higher sampling was found to be $>10^5\times$ slower for pixel counts as low as 512x512. The obtained results were subsequently confirmed with experimental demonstrations involving the effective reformatting of an SLM. Lastly, a time-averaging investigation demonstrated that NIMBLE-PATCH best mobilized the excess SLM refresh rate available for $N$-multiplexed operation as it achieved the best contrast while retaining a compute time advantage of several orders of magnitude relative to GS.

## 2. Results

### 2.1 Description of NIMBLE-PATCH algorithm and evaluation framework

We introduce the efficiency-driven point-to-patch assignment process of the NIMBLE-PATCH algorithm with a treatment of 3D point steering holograms and their target location-dependent variations in regional diffraction efficiency. For a given prototypical $2f$ CGH system with a Fourier-transforming lens of focal length $f$ and at an optical wavelength $\lambda$, the phase shift $\varphi$ required at $(x, y)$ positions across the hologram plane to achieve lateral point steering to a location $(x', y') = (d_{x'}, d_{y'})$ at the rear focal plane is given by:

$$\varphi_{lateral}(x, y) = -2\pi \cdot \left(\frac{d_{x'}}{\lambda f} x + \frac{d_{y'}}{\lambda f} y\right) \tag{1}$$

Similarly, the hologram phase mask required to achieve axial steering to a depth $z' = d_{z'}$ is a spherical profile paraxially approximated as a paraboloid [47] as given by:

$$\varphi_{axial}(x, y) = \frac{2\pi}{\lambda}\left(\frac{f^2}{d_{z'}} - \frac{f^2}{d_{z'}}\sqrt{1 - \left(\frac{d_{z'}}{f^2}\right)^2 (x^2 + y^2)}\right) \approx \frac{\pi d_{z'}}{\lambda f^2}(x^2 + y^2) \tag{2}$$

Convolving lateral deflection together with axial deflection for joint 3D point steering entails multiplying the complex axial and lateral fields together in the Fourier domain at the hologram plane, which corresponds to summing the phase profiles given in Eqs. (1) and (2):

$$\varphi_{3D}(x, y) = \frac{\pi}{\lambda f}\left(\frac{d_{z'}}{f} x^2 - 2d_{x'} x + \frac{d_{z'}}{f} y^2 - 2d_{y'} y\right)$$

$$= \frac{\pi d_{z'}}{\lambda f^2}\left(x - \frac{d_{x'} f}{d_{z'}}\right)^2 + \frac{\pi d_{z'}}{\lambda f^2}\left(y - \frac{d_{y'} f}{d_{z'}}\right)^2 - \frac{\pi}{\lambda d_{z'}}\left(d_{x'}^2 + d_{y'}^2\right)$$

$$\equiv \frac{\pi d_{z'}}{\lambda f^2}\left(x - \frac{d_{x'} f}{d_{z'}}\right)^2 + \frac{\pi d_{z'}}{\lambda f^2}\left(y - \frac{d_{y'} f}{d_{z'}}\right)^2 \tag{3}$$

Completing the square and dropping the piston phase offsets as shown in Eq. (3) demonstrates that 3D point-steering phase masks simply correspond to the parabolic profiles required for pure axial steering with a laterally shifted vertex location given by $(d_{x'} f / d_{z'}, d_{y'} f / d_{z'})$ [31,32]. Examples of such phase masks are illustrated in Fig. 1. As the target depth plane approaches the rear focal plane $d_{z'} = 0$, Eq. (3) simplifies to the pure lateral steering expression in Eq. (1), resulting in a uniform phase gradient as seen in Fig. 2(b).

Under the spatially discretized phase mask produced by an SLM of pixel pitch $p$ and pixel count $F \times F$, the regional diffraction efficiency $\eta$ at a given location $(x_0, y_0)$ on the SLM plane can be determined from the mean phase steps $\Delta\varphi_x(x_0, y_0)$ and $\Delta\varphi_y(x_0, y_0)$ to adjacent phase pixels at that location along the two orthogonal axes of the SLM as given by the following relationship [25]:

$$\eta(x_0, y_0) = \left(\frac{\sin\left(\frac{\Delta\varphi_x(x_0,y_0)}{2}\right)}{\frac{\Delta\varphi_x(x_0,y_0)}{2}}\right)^2 \left(\frac{\sin\left(\frac{\Delta\varphi_y(x_0,y_0)}{2}\right)}{\frac{\Delta\varphi_y(x_0,y_0)}{2}}\right)^2$$

$$= \left(\frac{\sin\left(\frac{p}{2}\frac{\partial\varphi}{\partial x}(x_0,y_0)\right)}{\frac{p}{2}\frac{\partial\varphi}{\partial x}(x_0,y_0)}\right)^2 \left(\frac{\sin\left(\frac{p}{2}\frac{\partial\varphi}{\partial y}(x_0,y_0)\right)}{\frac{p}{2}\frac{\partial\varphi}{\partial y}(x_0,y_0)}\right)^2 \tag{4}$$

Under purely lateral steering, i.e., $d_{z'} = 0$, the uniform phase gradient results in the following simplified relationship:

$$\eta(x_0, y_0) = \left(\frac{\sin\left(\frac{\pi p d_{x'}}{\lambda f}\right)}{\frac{\pi p d_{x'}}{\lambda f}}\right)^2 \left(\frac{\sin\left(\frac{\pi p d_{y'}}{\lambda f}\right)}{\frac{\pi p d_{y'}}{\lambda f}}\right)^2 \tag{5}$$

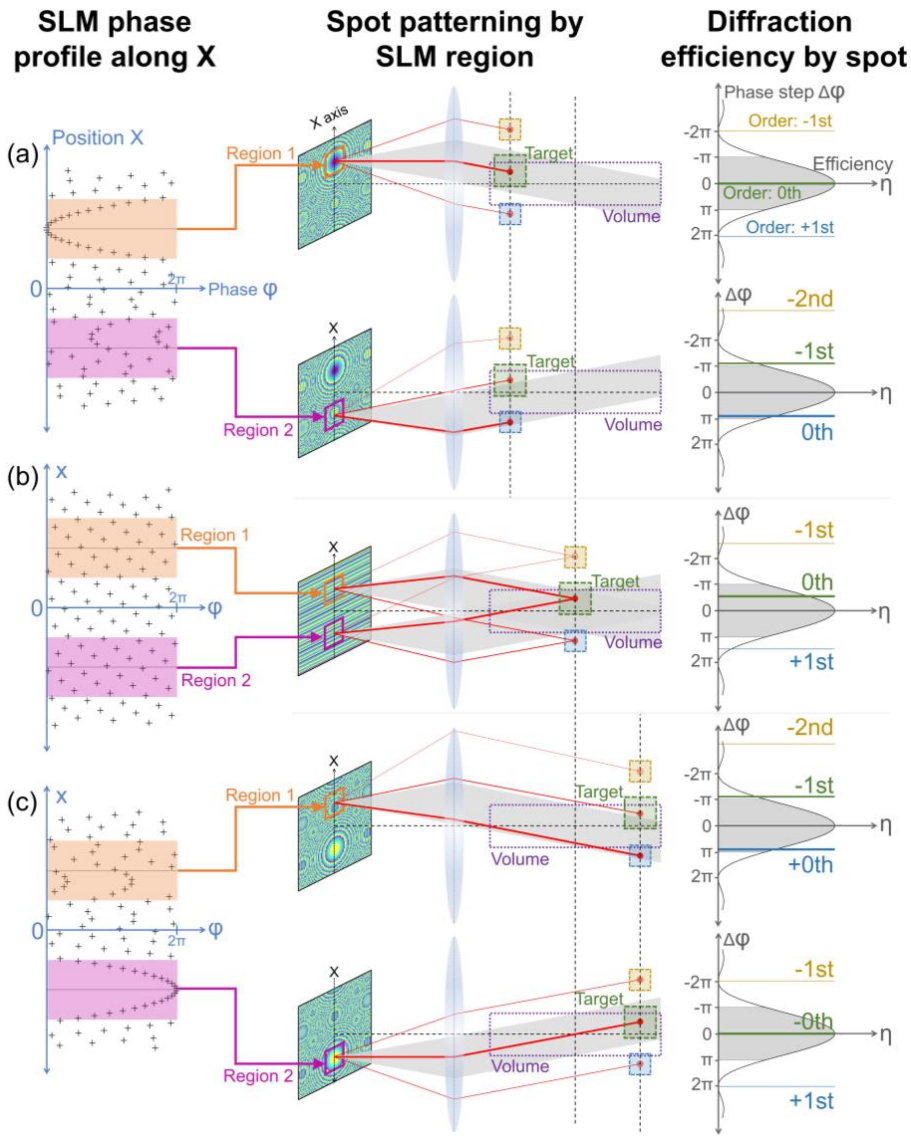

Fig. 2. Illustration of regional variations in phase mask efficiency under 3D point steering for a fixed lateral target position at (a) positive, (b) zero, and (c) negative target focus depths. For every aperture mask region applied to an SLM phase profile targeting a given location within the patterned volume's FoV, multiple diffraction spots are produced, one of which is the target. The optical power efficiency of each diffraction spot can be calculated from the corresponding unwrapped phase step $\Delta\varphi$ between adjacent SLM pixels at the center of the chosen region. For each region, the brightest diffraction spot is located within the angular steering range corresponding to the zeroth diffraction order (gray zone in patterning schematic) as the associated unwrapped phase gradient has the smallest magnitude such that $-\pi<\Delta\varphi<\pi$ (gray zone in efficiency plots). (a) Masking the SLM phase profile corresponding to the positive depth target position to span only Region 1, where the parabolic vertex is located, minimizes overall phase gradients such that target efficiency is maximized (denoted in green) and all other off-target diffraction orders (denoted in blue and yellow) have considerably lower efficiencies. Masking to Region 2, which is distant from the parabolic vertex, results in aliasing such that the target is effectively a higher-order diffraction spot (green): an off-target spot (blue) therefore becomes the brightest spot, and target efficiency is degraded significantly. (b) For the same lateral target position placed at zero depth, which corresponds to the rear focal plane of the Fourier-transforming lens, the phase gradient is uniform across the SLM and target efficiency is highest

among all diffraction orders as long as the target is in the lateral FoV. (c) For the same lateral target position at a negative depth, the parabolic vertex location is altered such that Region 2 becomes the peak target efficiency region while Region 1 patterns the target as a low-efficiency, higher-order diffraction spot.

Under joint axial and lateral steering, i.e., $d_{z'} \neq 0$, phase gradients along each axis scale linearly with distance to the parabolic vertex location $(d_{x'}f/d_{z'}, d_{y'}f/d_{z'})$ identified from Eq. (3), resulting in the following generalized expression for regional diffraction efficiency:

$$\eta(x_0, y_0) = \left(\frac{\sin\left(\frac{\pi p d_{z'}}{\lambda f^2}\left(x_0 - \frac{d_{x'}f}{d_{z'}}\right)\right)}{\frac{\pi p d_{z'}}{\lambda f^2}\left(x_0 - \frac{d_{x'}f}{d_{z'}}\right)}\right)^2 \left(\frac{\sin\left(\frac{\pi p d_{z'}}{\lambda f^2}\left(y_0 - \frac{d_{y'}f}{d_{z'}}\right)\right)}{\frac{\pi p d_{z'}}{\lambda f^2}\left(y_0 - \frac{d_{y'}f}{d_{z'}}\right)}\right)^2 \quad (6)$$

The expression in Eq. (6) can serve to evaluate the relative contributions of different regions of the SLM to the targeted spot under 3D point steering. As illustrated in Fig. 2, the most efficient SLM region for a given target corresponds to its phase mask's parabolic vertex, whose location depends on the 3D position of the targeted spot. Additionally, once the phase step $\Delta\varphi(x_0, y_0)$ at a given SLM location away from the vertex exceeds $\pi$ along either axis, the targeted spot falls outside of the achievable angular diffraction range of $arctan(\lambda/p)$ at that SLM location. Alternatively stated, that SLM region directs the bulk of its optical power to an off-target 3D point position considered to be within the region's zeroth diffraction order, thereby contributing only marginal power to the targeted 3D point position as a higher diffraction order. This behavior has implications on sampling requirements for FFT-based CGH computation approaches, including GS. The least computationally burdensome implementations of such approaches involving minimal sampling of the hologram plane (i.e ~1 computational pixel per SLM pixel) risk poor performance by failing to properly account for relative contributions to different diffraction orders as result of aliasing [48]. This impact is especially prominent with increased axial steering away from $z' = 0$, which entails steeper parabolic phase gradients.

Given a 3D point cloud consisting of multiple target point positions with different associated peak-contribution regions across the SLM plane, NIMBLE-PATCH exploits the described deterministic relationships to partition the SLM across points, separately allocating SLM pixels to each target location for maximum overall efficiency. Under the NIMBLE-PATCH implementation covered in this work, the paraxial spot size requirement is uniform across all targets such that patches are evenly sized and arranged in a tiled geometry. The allocation of each target to each patch is accomplished by calculating regional diffraction efficiencies at each patch center for each target using Eqs. (5) and (6) in order to construct a cost matrix. The associated linear assignment problem between each distinct patch in every available SLM frame and each separate target spot is solved via the Jonker-Volgenant method [49]. Since the objective of the assignment was chosen to be the maximization of overall diffraction efficiency, if a given patch results in optimally high efficiency across several proximal target locations, precedence is given to the target with the highest absolute efficiency. Finally, phase masks are computed separately for each target and only across the corresponding SLM patch using the simple beamsteering and focus tuning relationships in Eqs. (1) and (3), then stitched together to generate the full-frame phase profiles.

Since CGH computation algorithm performance is heavily dependent on SLM pixel count $F \times F$ and targeting throughput, given by total target count $T$ across $N$ time-multiplexed frames, NIMBLE-PATCH was implemented and evaluated against GS across sweeps involving all three parameters, as shown in Code 1 [50]. Two versions of GS were included in the comparison: one implementation denoted by GSx1 minimizes computation with a sampling scheme of 1x1 computational pixel per SLM pixel, and a second implementation denoted by

GSx3 prioritizes performance at the expense of computation burden with a sampling scheme of 3x3 computational pixels per SLM pixel. In order to ensure that the comparison across algorithms is agnostic to context-dependent optical system parameters (including optical wavelength, lens focal length, and SLM pitch), sweeps across $F$, $T$, and $N$ were normalized to the achievable resolution and FoV of the unit SLM patch targeting a distinct point. Specifically, lateral and axial FoVs were confined to the allowable volume determined by the patch format and target spot sizes used for irradiance, contrast, accuracy, and efficiency calculations were set based on the patch's achievable lateral resolution. The impact of NIMBLE-PATCH partitioning on target FoV and spot dimensions is illustrated in Fig. 3.

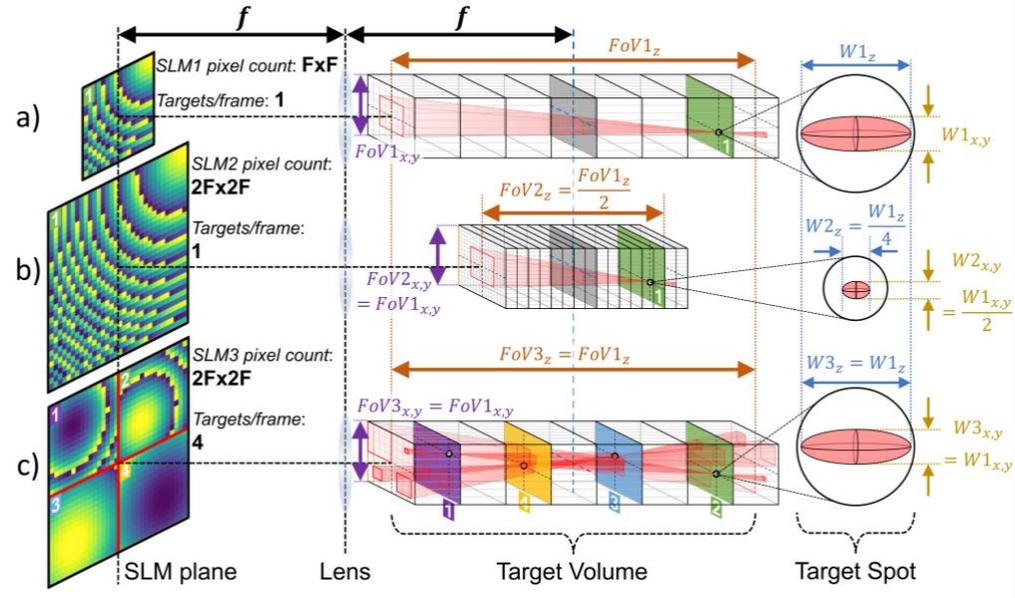

Fig. 3. SLM partitioning principle underlying the NIMBLE-PATCH algorithm with SLM pixel count and phase mask on the left, volume dimensions and target point cloud in the center, and target spot dimensions on the right. Both field-of-view and spot size dimensions are determined by the partitioned patch size for a given *2f* system. (a) Case 1: A system consisting of an SLM (denoted as *SLM1*) of pixel count *FxF* targeting 1 spot/frame and a Fourier-transforming lens of focal length *f* results in an accessible target volume spanning a lateral dimension $FoV1_{x,y}$ and an axial dimension $FoV1_z$, as well as a patterned spot of lateral size $W1_{x,y}$ and axial size $W1_z$. (b) Case 2: Doubling the SLM size along each dimension under fixed pitch for an *SLM2* pixel count total of *2Fx2F* under the same lens focal length *f* and targeting throughput of 1 spot/frame maintains the same lateral field-of-view as Case 1 ($FoV2_{x,y}=FoV1_{x,y}$), but halves the axial field-of-view to $FoV2_z=FoV1_z/2$, halves the lateral spot size to $W2_{x,y}=W1_{x,y}/2$, and divides axial spot size by 4 to $W2_z=W1_z/4$. (c) Case 3: Subsequently increasing targeting to 4 spots/frame for the same system as Case 2 reduces the patch size in *SLM3* to the *FxF* pixel count of *SLM1*, resulting in the same FoV and spot size dimensions as Case 1.

As evidenced by Eq. (5), lateral point steering is subject to a gradual efficiency roll-off that caps lateral FoV as given by $FoV_{x,y} = \lambda f/p$. The $F$-fold increase of this FoV relative to the Abbe diffraction limit for lateral spot width $W_{x,y} = (\lambda f)/(pF)$ captures the degrees of freedom available for lateral steering along one axis. An axial FoV can similarly be calculated by noting that lateral FoV is capped at $\Delta \varphi = \pi$ and solving for the depth at which mean $\Delta \varphi$ along a given SLM axis is equal to $\pi$: $FoV_z = (16f^2\lambda)/(F(\lambda^2 + 4p^2)) \approx (4f^2\lambda)/(Fp^2)$. The $F/2$-fold increase of the axial FoV relative to the Abbe diffraction limit for axial spot width $W_z =$

$(8f^2\lambda)/(F^2p^2)$ can be attributed to the fact that focus tuning employs circularly symmetric phase masks with a radius spanning $F/2$ pixels along either SLM axis.

Efficiency roll-off profiles along both a single lateral axis and the axial dimension are shown in Fig. S1, with axial steering experiencing compounded loss relative to lateral 1D steering as a result of the 2D radial phase modulation required for focus tuning. These efficiency roll-offs may place additional constraints on allowable FoV for applications that require minimal variation in spot power across point clouds [51]. For instance, optical interfacing in biological microscopy sets a minimum bound on spot power based on the required photoexcitation as well as a maximum bound based on tissue heating and photobleaching limits [3,22]. Accordingly, lateral and axial FoV ratio parameters were incorporated into our implemented simulation framework in order to provide the option of constraining 3D point cloud volume to desired fractions of the full FoVs. We note that an axial range ratio of 0.25 can serve to ensure that on-axis targets (i.e. targets along $z$ at $(x', y') = (0,0)$) remain within the zeroth diffraction order across all regions of the SLM for FFT-based algorithms involving minimal sampling [16,35]. However, this restriction remains insufficient for off-axis points ($(x', y') \neq (0,0)$), such that costlier concessions would be needed to ensure zeroth order targeting across the full point cloud volume.

*2.2 Single-frame algorithm performance comparison*

NIMBLE-PATCH was first compared against GSx1 and GSx3 under single-frame patterning ($N = 1$ frame) for randomly distributed 3D point clouds. The comparison was made across 5 SLM pixel counts ($F \times F = 32 \times 32, 64 \times 64, 128 \times 128, 256 \times 256, 512 \times 512$) and up to 5 total target counts ($T = 1, 4, 16, 64, 256$ targets). Lateral and axial range ratios were set to 0.9 and 0.75, respectively, for the target volume. This reduction in FoV ensures that all generated spots have a theoretical efficiency >7.5% (Fig. S1). In order to accommodate the discretization limitation of the GS algorithms, allowable target point depths were constrained to a finite number of depth planes evenly spaced across the axial FoV such that the separation between depth planes was equal to the axial spot size. Maximum target count simulated for a given $F$ is therefore reached once distinct, non-overlapping spots can no longer be placed across these target depth planes. For each $(F, T)$ pair, at least 25 randomized distributions were simulated (>80 for $F < 512$), and each randomly generated distribution was evaluated across all three algorithms.

For each simulation, the target volume intensity distribution is denoted with $I(x', y', z')$ and the generated volume intensity distribution is denoted with $G(x', y', z')$. $I(x', y', z')$ is discretized in $z$, and is only defined for depth planes that include targets. Targets in $I(x', y', z')$ are disk-shaped pixel regions with a value of 1 sized to the achievable resolution of the corresponding patch size under NIMBLE-PATCH, and $I(x', y', z')$ is 0 across all non-target regions. Consequently, both algorithms are used to generate identically sized spots across the same FoV. In order to quantify the results of the simulations and compare CGH algorithms, two main metrics were considered: contrast $C$ and computation time. Contrast measures the ratio of irradiance in target regions to irradiance in non-target regions:

$$C = \frac{\frac{\sum G(x',y',z')\,I(x',y',z')}{\sum I(x',y',z')}}{\frac{\sum G(x',y',z')\,(1-I(x',y',z'))}{\sum (1-I(x',y',z'))}} \quad (7)$$

Computation time measures the time elapsed between the input of the volume information to the algorithm (point cloud target coordinates for NIMBLE-PATCH, $I(x', y', z')$ for GSx1 and GSx3) and the output of the computed phase mask. Reported computation times were obtained from phase mask calculations performed on a single core of the Intel Xeon E5-2670 v3 CPU and 384 GB of RAM.

In addition to contrast and computation time, two supplementary metrics were calculated for each distribution: *accuracy* $\alpha$ and *efficiency* $\eta$ (see Supplement 1). The choice to present contrast as the main metric stems from its importance in point-cloud holography applications [3] and its well-behaved and predictable scaling behavior as a function of target count. As target count quadruples, the SLM patch area dedicated to each target shrinks by a factor of 4, reducing the available optical power directed to the spot by the same factor. Furthermore, since the patch size reduction also corresponds to an increase in spot size by the same factor as depicted in Fig. 3, target irradiance drops by an additional factor of 4. This results in a theoretical decrease in target irradiance that is proportional to the square of the number of targets, i.e., a ~40 dB/decade decay in contrast plotted against the target count (ignoring the associated increase in the background irradiance and without taking algorithm-specific performance into account).

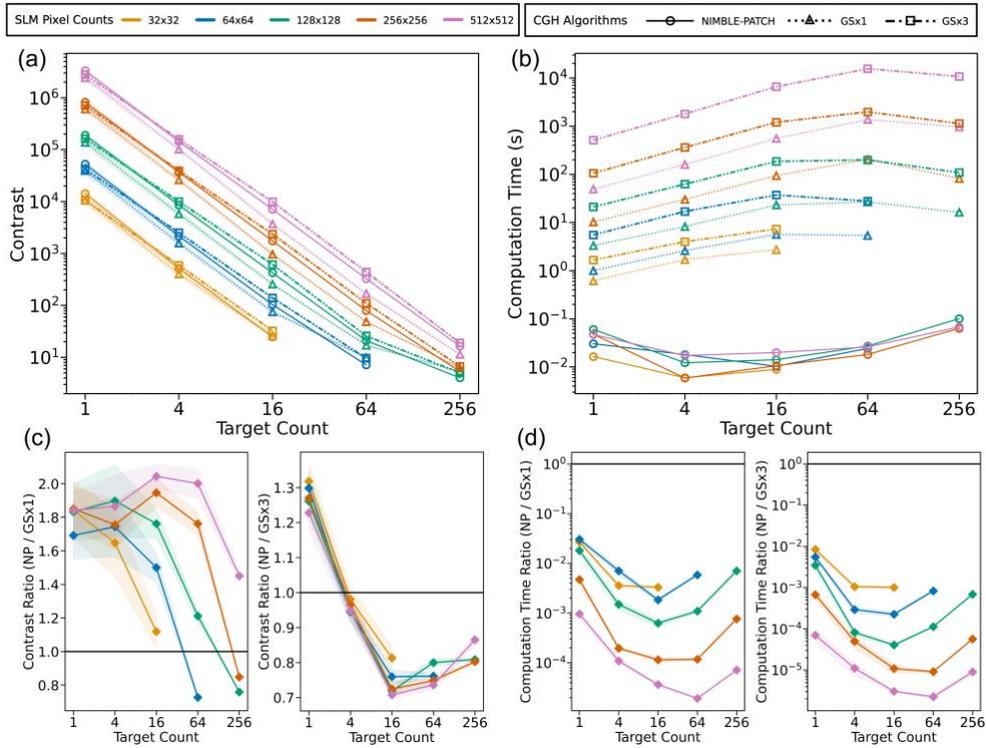

Fig. 4. Sweep results comparing NIMBLE-PATCH to GS for single-frame simulations. (a) Contrast and (b) computation time as a function of target count $T$ across SLM pixel counts $F \times F$. (c) Contrast ratios and (d) computation time ratios between NIMBLE-PATCH and the two GS algorithms (top: GSx1, bottom: GSx3). 95% confidence intervals for the mean value of the curves are shown as shaded regions for each plot.

Fig. 4 (a) and (b) show aggregated contrast and computation time results, respectively, across all SLM pixel counts from single-frame simulation runs (see Fig. S2 for $\alpha$ and $\eta$). Contrast follows the general 40 dB/decade decay trend for all algorithms across all pixel counts, with the added contribution of algorithm-specific performance deviations. For NIMBLE-PATCH, computation time initially drops with increasing target count $T$ due to the speed-up of phase mask computations under increased SLM partitioning resulting in smaller patch sizes. Computation time then rises, converging across pixel counts $F \times F$, as the linear sum assignment solution, which depends only on $T$, emerges as the limiting factor. For GSx1 and

GSx3, increasing $T$ raises computation time on average as the number of distinct target depth planes that need to be accounted for grows together with point cloud density. However, once $T$ exceeds a certain threshold for each $F$, a roll-off in computation time occurs as randomly generated target locations reliably cover all allowable target depths. Any subsequent increase in $T$ causes a decrease in this cap of allowable depth planes, speeding up computation. Fig. 4(c) shows the ratios between contrast in NIMBLE-PATCH and contrast in either of the two implementations of GS, while Fig. 4(d) shows the corresponding computation time ratios. NIMBLE-PATCH consistently outperforms GSx1 both in contrast and computation time. Up until absolute contrast values drop below ~10, NIMBLE-PATCH achieves on average ~1.6-1.8x higher contrast compared to GSx1, likely as a result of aliasing from minimal sampling in GSx1 [48]. NIMBLE-PATCH was also found to complete phase computation 1.5 – 5 orders of magnitude faster than GSx1 for the considered SLM pixel counts. GSx3, on the other hand, matches NIMBLE-PATCH patch performance as early as $T > 1$ at the expense of increased computation, further corroborating the sampling-related limitations of GSx1. Therefore, the computation time ratio between NIMBLE-PATCH and GSx3 is even more considerable, reaching 2 – 6 orders of magnitude. Given that the relative computational burden of GS rises consistently with increasing SLM pixel count, this yawning gap in computation time ratio can be expected to increase beyond 5 – 6 orders of magnitude at SLM pixel counts exceeding 512x512.

*2.3 Experimental evaluation of algorithm performance*

Algorithm performance findings from simulated sweeps were verified in experiment under the imaging setup shown in Fig. 5(a) using a 532 nm diode-pumped solid-state laser source (Thorlabs CPS532). Phase masks computed by each algorithm across a set of target point clouds were applied using a 1272x1024 SLM (X13138-01 12.5µm-pitch Hamamatsu LCoS) and the resulting patterned volumes were acquired as z-stacks using a 12-bit camera (BlackFly S BFS-U3-200S6M-C) mounted onto an automated z-stage (Zaber X-LSQ150A). In order to replicate the sweeps performed in simulation across SLM pixel count $F \times F$ and target count $T$, the central 1024x1024 pixel region of the SLM was reformatted to smaller effective pixel counts via real pixel grouping. A binary 0 and $\pi$ checkerboard phase pattern was applied at a pitch of 2 real pixels across the remaining regions of the rectangular-format SLM in order to diffract light away from the point cloud volumes patterned by the central reformatted square SLM area. The range of allowable values for $F$ and $T$ was jointly constrained by this reformatting scheme, the largest axial FoV that can be accommodated by the stage, and the sampling limit set by the pixel pitch of the camera. In order to better accommodate $(F, T)$ sweeps, a focal length of 100 mm was chosen for the Fourier-transforming lens, and the lateral and axial range ratios constraining targeted point cloud volumes were chosen to be 0.8 and 0.5, respectively. Three effective SLM pixel counts ($F \times F = 32 \times 32, 64 \times 64, 128 \times 128$) and three target counts (T=1, 4, 16) were investigated, with two point-cloud distributions evaluated across all three algorithms for each $(F, T)$ pair, for a total of 54 unique phase masks patterning distinct volumes.

Representative images of computed phase masks and 2D maximum intensity projections of both simulated and acquired point cloud volumes are shown in Fig. 5(b-f). The corresponding axial sweep videos for the simulated and acquired volumes are also provided in Visualizations 1, 2, 3, 4, and 5. The obtained projections show close agreement between simulated and measured results. Experimental non-idealities include some depth distortion from target depth-dependent aberrations (see Supplement 1), and a central spot formed at $(x', y', z') = (0,0,0)$ from the unmodulated illumination and DC diffraction component resulting from the physical SLM's finite pixel fill factor, phase error, etc. GSx1 performance was found to be consistently worse than NIMBLE-PATCH, with a reduction in target power that is increasingly apparent at higher target counts and distant depth planes (Fig. 5(e-f)). In addition, the disappearance of certain target spots at smaller target counts (Fig. 5(b,d)) further confirms the susceptibility of the minimally sampled GSx1 algorithm to aliasing impacts. Qualitatively, GSx3 performance was found to be comparable to that of NIMBLE-PATCH. We note that both GS algorithms tend to produce targets with higher peak intensities at the infinitesimal target spot centers, despite having worse or comparable integrated optical power across target spot sizes relative to NIMBLE-PATCH. This can be attributed to the fact that patch partitioning is not enforced in GS. GS pixel ensembles targeting separate point locations are thus interleaved across the SLM, leading to higher spot resolution, as revealed by an examination of generated phase masks in Fig. 5(e-f).

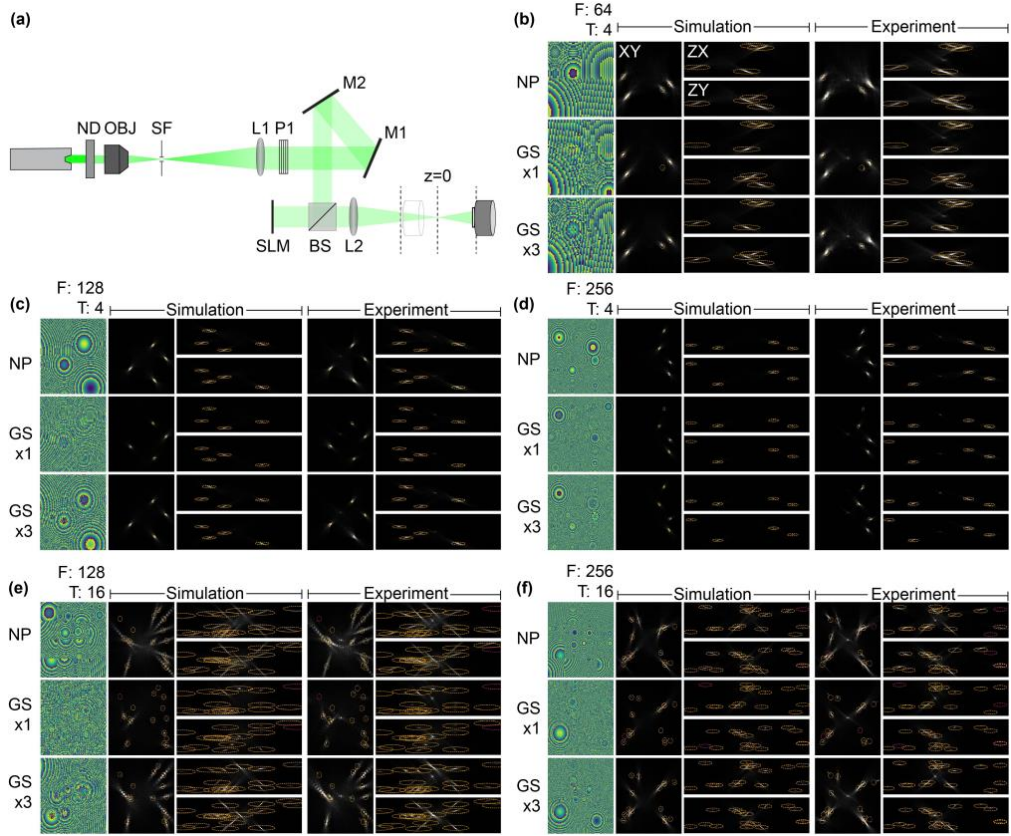

Fig. 5. (a) Experimental test setup for point-cloud volume acquisitions under SLM reformatting. ND: neutral density filter, OBJ: objective (10x), SF: spatial filter (25 μm pinhole), L1: collimating lens (focal length: 180 mm), P1: polarizer, M1/M2: alignment mirrors, BS: beamsplitter, L2: Fourier-transforming lens (focal length: 100 mm). Representative images of phase masks computed by each algorithm along with simulated and experimental 2D maximum intensity projections for a range of pixel count $F \times F$ and target count $T$ conditions: (b) $F \times F = 64 \times 64$, $T = 4$ (see Visualization 1), (c) $F \times F = 128 \times 128$, $T = 4$ (see Visualization 2), (d) $F \times F = 256 \times 256$, $T = 4$ (see Visualization 3), (e) $F \times F = 128 \times 128$, $T = 16$ (see Visualization 4), and (f) $F \times F = 256 \times 256$, $T = 16$ (see Visualization 5). Projection images span the FoV associated with each (F,T) condition, projection orientations across image layouts are specified in (b). Identified and non-identified targets are marked by orange and red ellipses, respectively.

Algorithm performance was quantified through a processing pipeline that identifies peak intensity locations and assigns them to target positions based on proximity (additional results and details available in the Supplement 1). Target lateral full widths at half maximum

(FWHMs) generally show good agreement between measured, simulated, and expected spot sizes (see Fig. 6(a)), with some deviation in FWHM estimation emerging in experiment from either proximity to the static central spot (e.g. at $F \times F = 64 \times 64, T = 1$) or a large FoV that is more susceptible to aberration impacts (e.g. at 256x256 formats where axial FoVs reach 70 mm). Spot size was also found to be consistently smaller with GS algorithms, confirming that effective single-target SLM patch regions in GS are larger than those in NIMBLE-PATCH from phase mask interleaving. The accuracy of the simulation framework is further evidenced by the fact that experimentally measured target irradiances recapitulate the same trends seen in simulation, as shown in Fig. 6(b). Importantly, we note that irradiances in NIMBLE-PATCH and GSx3 follow each other closely, indicating that both algorithms perform similarly in their ability to allocate power to each target location. In alignment with simulation sweep results and qualitative evaluations of experimental projections, markedly degraded irradiance performance is also observed with GSx1.

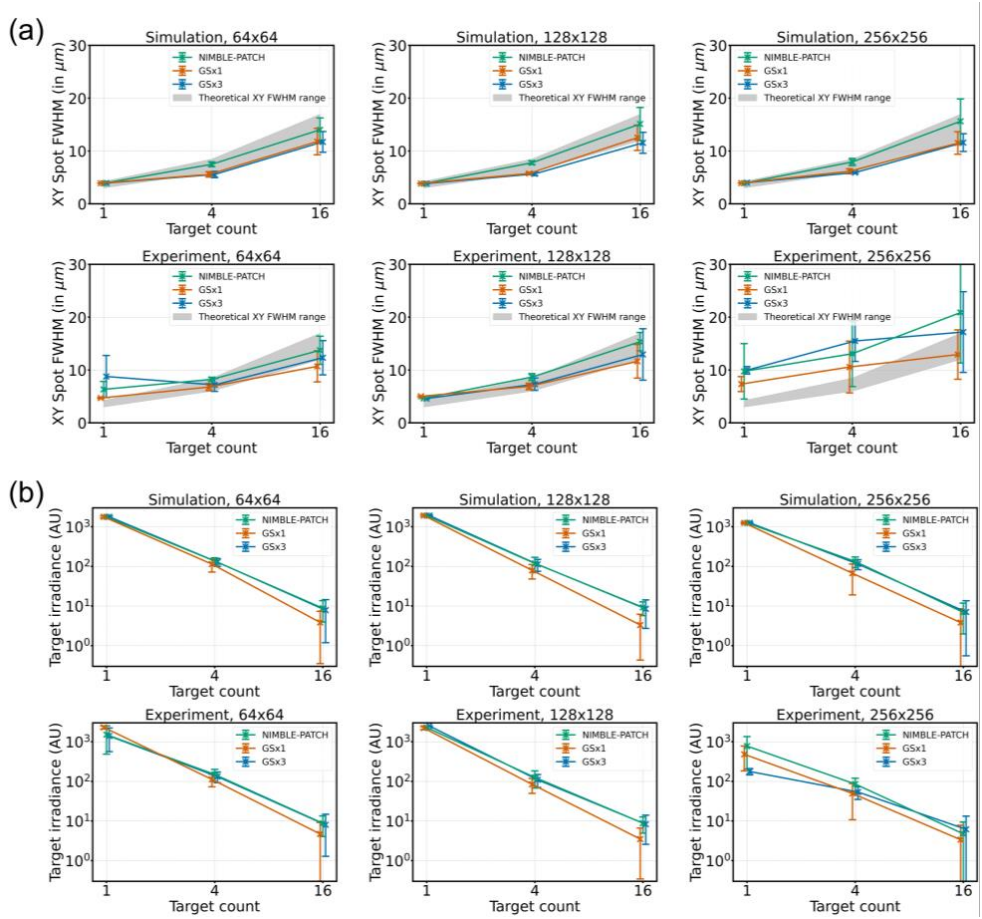

Fig. 6. (a) Lateral FWHM spot sizes measured across simulated and experimental point cloud volumes evaluated at each pixel count $F \times F$ and target count $T$ condition (error bars: standard deviation). Shaded gray region, spanning theoretical FWHMs corresponding to the side and diagonal apertures of associated square SLM patch, is shown as a reference range due to spot sampling and pixel integration limitations with the simulation environment and camera pitch. (b) Target irradiances in arbitrary units across simulated and experimental point cloud volumes evaluated at each $(F, T)$ condition (error bars: standard deviation).

*2.4 Time multiplexing performance evaluation*

In order to maximize the number of distinctly resolvable target positions that can possibly be accessed in the target volume, the target count per frame (and hence patch count in NIMBLE-PATCH) must be minimized. However, the number of targets that must be generated within a given timeframe may be dictated by application-specific requirements. If the SLM has excess refresh rate capable of switching between multiple holograms above the temporal bandwidth relevant to the application, time averaging across multiple holograms can be employed to simultaneously achieve the desired resolution and target count across the given FoV and timeframe [19,42,43,52]. NIMBLE-PATCH can be time-multiplexed since the total set of $T$ targets are optimally distributed across frame-specific patches as part of the assignment process. The same efficiency-driven inter-frame target decomposition process can be applied to GS, with each individual phase mask being computed by GS instead of NIMBLE-PATCH after points are assigned to separate frames. For each 3D point cloud, a single GS hologram addressing the complete set of targets within just one frame was also computed to serve as a benchmark. The decomposed NIMBLE-PATCH, decomposed GSx1, and single-frame GSx1 benchmark algorithms were evaluated for their relative abilities in making optimal use of time-multiplexing capabilities via performance sweeps across available excess refresh rate. GSx3 was not considered in this analysis as it could not be viably investigated for time-multiplexed performance given its significant computational burden.

Fig. 7 shows the results of excess refresh rate ($N$) sweeps at two total target count values ($T = 64$ and $T = 256$). As $N$ increases, the target count per frame (and therefore the patch count in each NIMBLE-PATCH hologram) decreases, which results in a decrease in spot size and an increase in addressable depth plane count. At this higher spatial degree-of-freedom regime, contrast generally increases due to shrinking spot size relative to the accessible FoV. Notably, decomposed NIMBLE-PATCH outperforms both decomposed GS and single-frame GS in higher $N$ regimes, with a crossover point that depends on $F$ (but occurs within $N < 16$ regime for all cases). As increasing $N$ decreases patch count per frame, the complexity of the balanced assignment problem remains constant in NIMBLE-PATCH, resulting in a relatively flat computation time across the refresh rate space. Another observation from these sweeps is that GS does not benefit from excess refresh rate in the same way NIMBLE-PATCH does, as single-frame GS outperforms decomposed GS for most cases.

Fig. 7 also serves to evaluate general tradeoffs between SLM pixel count and speed. Spot size and FoV can simultaneously be made equal across two different SLM pixel counts $F_1 \times F_1$ and $F_2 \times F_2$ for a given $T$ by choosing $N_1/N_2 = (F_2/F_1)^2$. For most cases, higher $N$ and smaller $F$ SLMs perform better, pointing to the benefit of utilizing a faster SLM in point-cloud targeting applications. However, as target-count-per-frame ($T/N$) approaches 1 (i.e., point-scanning regime), a drop in contrast relative to lower $N$ and higher $F$ devices is observed. This is due to the fact that partitioning is obviated when $T/N = 1$, eliminating the advantage of the optimized assignment process that places patches nearer to the parabolic vertex locations of their corresponding targets for improved efficiency.

## 3. Discussion and conclusions

NIMBLE-PATCH features sub-100-ms computation times across nearly all evaluated single-frame cases. It therefore already matches the fastest existing GPU-based CGH algorithms, all without utilizing parallelization, without making concessions in volumetric point-cloud quality (e.g., severely constraining addressable depth plane count), and without the need for time-intensive and context-specific model pre-training [37,38]. Importantly, NIMBLE-PATCH boasts a highly-parallelizable architecture where SLM pixel values are independent of one another and computed non-iteratively in a manner that is compatible with single instruction multiple data (SIMD) processing. These features allow for NIMBLE-PATCH to be scaled both to low-latency computation through the use of parallelized hardware implementations (GPU or FPGA-based) for real-time phase retrieval in closed-loop applications, and to hardware-light computation on resources like single-core CPUs or dedicated lightweight DSP pipelines implementing Eq (3) on a per-pixel basis. To demonstrate the scalability of NIMBLE-PATCH in even consumer-grade CPU-based environments, we tested SLM pixel counts of $F = 1024 \times 1024, 2048 \times 2048, 4096 \times 4096$ across target counts of $1 - 256$, on an Apple M1 Pro CPU and 16 GB of RAM, recording compute times ranging between 100 ms and 1 s. These same SLM phase mask formats cannot be feasibly computed on CPU by CNN- or FFT-based algorithms like GS, all of which require GPU-based computation at larger formats.

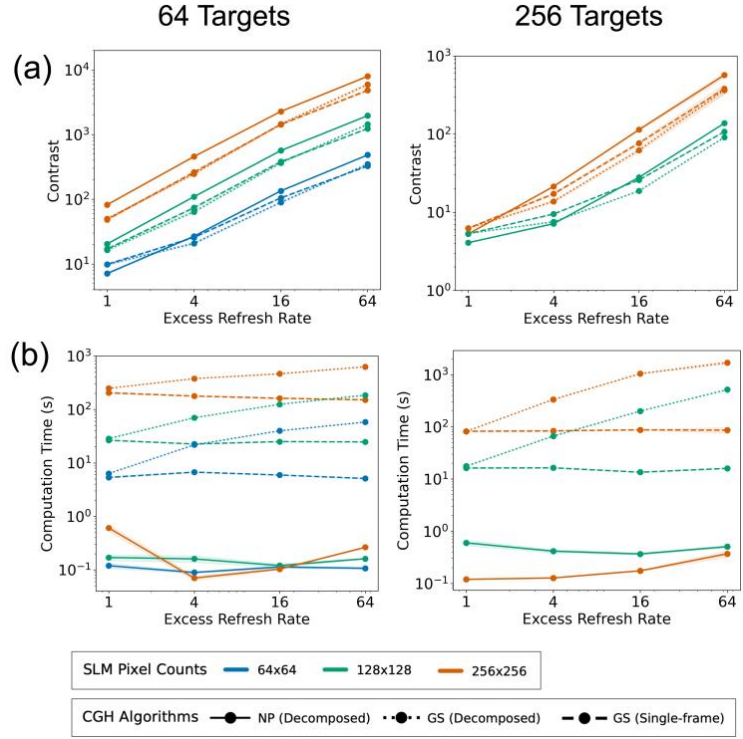

Fig. 7. Sweep results comparing NIMBLE-PATCH to decomposed GS for excess refresh rate scaling evaluation, using single-frame GS as a benchmark. (a) Contrast and (b) computation time scaling as a function of available excess refresh rate $N$ across SLM pixel counts $F \times F$, for two total target count $T$ scenarios: $T = 64$ and $T = 256$. 95% confidence intervals for the mean value of the curves are shown as shaded regions for each plot.

Though the NIMBLE-PATCH implementation evaluated in the work was specifically designed to simultaneously exploit the full FoV allowed by the SLM pitch and maximize overall efficiency (and therefore contrast), an entire class of implementations can be envisioned using the same underlying principles. For instance, maximizing overall efficiency may lead to the deprioritization of some distant targets with inherently low efficiency, impacting spot power uniformity. Accordingly, alternative cost functions or heuristics may be employed instead of the efficiency-based linear sum assignment in order to prioritize the weakest performing spots by assigning them to their optimal patches first. Similarly, target-specific spot irradiance constraints may be imposed by calculating optimal distances to parabolic vertices for each target based on desired efficiency, then minimizing deviation from these distances instead of distance to the vertex itself during patch assignment. Control over relative spot power across point cloud volumes can further be aided by redundantly assigning a different number of patches across different frames to each target, which also offers flexibility on allowable target count. Patch shapes may be altered to allow for some amount of interleaving as well, such that high-desirability patches can be shared across multiple targets and spot resolution can be increased (at the expense of FoV). Furthermore, NIMBLE-PATCH is capable of producing cloud volumes from a variety of primitives instead of diffraction-limited spots, including lines and polygons, since convolving a desired shape across a point-cloud simply requires adding the required phase mask to each patch [53,54]. The partitioning geometry decomposing the SLM into patches may additionally be altered to better accommodate such primitives (e.g. strip-shaped patches for line primitives). Lastly, time multiplexed operation may be mobilized to improve speckle instead of throughput via random phase offsets applied to each patch in order to average out interference between distinct targets [42,44,45].

The widening gap between NIMBLE-PATCH and GS computation times observed under increasing SLM pixel count in this work indicates that, at state-of-the-art SLM pixel counts beyond 512x512 and under the same compute resources, NIMBLE-PATCH can be expected to operate >10,000x faster than lightweight iterative CGH algorithms and >100,000x faster than iterative CGH algorithms of matched performance [25,26]. This speedup is obtained by employing a predetermined scheme to apportion the SLM's available pixels across targets, thereby circumventing the computational burden of full-frame computation for each additional target. However, we note that this considerable speed advantage is earned at the expense of patterning flexibility as layer-based CGH phase masks can make use of interleaving or shared SLM pixels across multiple targets to account for mutual interference between targets. Such functionality may be more suitable for low-contrast volumes such as dense point clouds or continuous 3D intensity distributions. Layer-based algorithms, including GS, could especially be more appealing with a reduced number of discrete planes since their computation is heavily dependent on depth plane count. However, given the multiple orders of magnitude separating GS and NIMBLE-PATCH computation times, a drastically limited number of depth planes would be required for GS to be reasonably competitive with NIMBLE-PATCH. Similarly, CNNs must limit their depth plane counts to ~10 in order to achieve computation times on the order of 10-100 ms [37,38]. Therefore, in the context of sparse point-cloud patterning, the NIMBLE-PATCH approach uncovers a new computation regime that simultaneously offers a path to real-time 3D point cloud holography, the ability to place targets with infinitesimal precision across the maximum FoV allowed by the SLM, and scaling behavior aligns with emerging time multiplexing methods under ongoing SLM improvements [26,47,55,56].

## 4. Back matter

**Funding.** National Science Foundation (2146752); McKnight Foundation (050023); National Institutes of Health (1RF1NS128772-01).

**Disclosures.** NTE, CY, RM: The Regents of the University of California (P).

**Data availability.** Implemented code and generated data presented in this paper are available in Code 1, Ref. [50].

**Supplemental document.** See Supplement 1, Code 1, Visualization 1, Visualization 2, Visualization 3, Visualization 4, and Visualization 5 for supporting content.

# FAST, NON-ITERATIVE ALGORITHM FOR 3D POINT-CLOUD HOLOGRAPHY: SUPPLEMENTAL DOCUMENT

## 1. Simulation framework

The simulation framework constructed for this paper (implemented in Python 3.9) models SLM-driven volumetric point cloud patterning under a $2f$ optical configuration involving a Fourier-transforming lens via Fast-Fourier Transform (FFT) and Fourier-domain Fresnel propagations. For GS computation, an iteration count of 50 was used to ensure a sufficient level of convergence for the point cloud hologram [1,2]. Each SLM pixel is represented by a programmable number of simulation pixels (set to 1x1 for GSx1 phase mask computation, 3x3 for GSx3 phase mask computation, and 5x5 for overall algorithm evaluation and metric calculation) to capture higher diffraction orders in the target volume. Additionally, the SLM phase mask is zero-padded (with a pad size equal to SLM size on all sides) to position diffraction-limited target spots with sub-diffraction-limit precision, to represent the spot shape as a disk for improved GS performance, and to minimize binning-related inaccuracies from FFT operations due to insufficient spatial granularity. To account for electronic drive non-idealities, each computed phase mask is discretized to a programmable number of bits (set to 8 in all performed sweeps). Efficiency roll-off profiles obtained from the simulation framework along both a single lateral axis and the axial dimension are shown in Fig. S1.

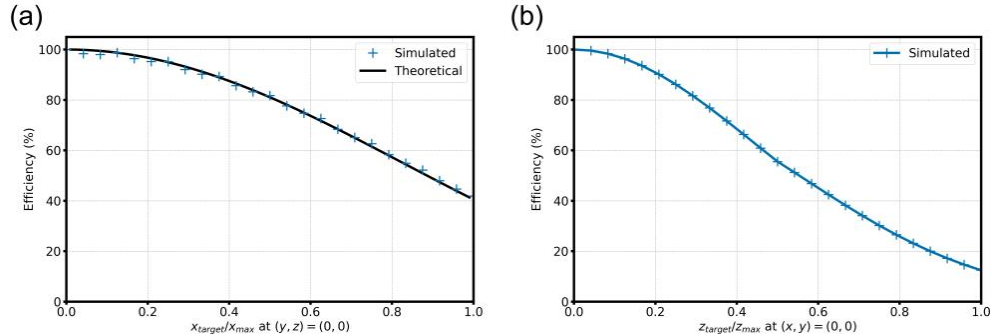

Fig. S1. Efficiency roll-off profiles along the (a) X axis and (b) Z axis obtained from single point steering using NIMBLE-PATCH in the simulation framework. Lateral 1D steering aligns with theoretical prediction, and axial steering experiences compounded loss relative to lateral 1D steering as a result of the 2D radial phase modulation required for focus tuning.

Axial range and depth plane count for random target generation were computed from each run's $F$, $T$, and $N$ values. Target location generation was further constrained in the axial dimension to a discrete set of distinctly resolvable depth planes for the computational benefit of GS, as each additional depth plane linearly impacts GS computation while not changing the computational complexity of NIMBLE-PATCH. In practice, spot positioning may require sub-spot-size precision in applications including super-resolution microscopy, further increasing the computational burden required of GS [3]. In the lateral dimensions, target positions were generated randomly across any of the allowed depth planes, with placement granularity determined by the choice of zero padding. An additional target spacing constraint was introduced to prevent target spots from overlapping.

To evaluate and compare the generated volume intensity distribution, two metrics were considered in addition to contrast and computation time: accuracy $\alpha$ and efficiency $\eta$. Accuracy

$\alpha$ corresponds to the cross-correlation between desired and generated intensity distributions and is defined as:

$$\alpha = \frac{\sum G(x',y',z')\, I(x',y',z')}{\sqrt{(\sum G(x',y',z')^2)\, (\sum I(x',y',z')^2)}} \qquad (S1)$$

Efficiency $\eta$ corresponds to the ratio of the power in the targets to the total power in the volume and is defined as:

$$\eta = \frac{\sum G(x',y',z')\, I(x',y',z')}{\sum G(x',y',z')} \qquad (S2)$$

Fig. S2(a,b) shows single-frame scaling trends for these two metrics across SLM pixel count and target count per frame, and Fig. S2(c,d) provides ratiometric comparison results between the three algorithms. NIMBLE-PATCH outperforms GSx1 in terms of accuracy $\alpha$ across most $F$ and $T$ cases, and GSx3 at low $T$ regimes in all $F$ cases. It is important to note that efficiency $\eta$ is closely related to contrast $C$. For a given target distribution, efficiency $\eta$ can be approximated to be contrast $C$ scaled by the ratio of total target spot area to the total size of the non-target area across all depth planes:

$$\eta \approx \frac{\sum I(x',y',z')}{\sum 1 - I(x',y',z')} C \qquad (S3)$$

This approximation holds for sparsely populated target point cloud volumes, which is generally applicable across the performed sweeps. Though the scaling factor depends on the targeted point cloud, it remains equal across algorithms for a given set of targets. This relationship explains why efficiency ratios, shown in Fig S2(d), closely follow the contrast ratios previously shown in Fig 4(c). The motivation behind presenting contrast $C$ as the main metric stems from its implicit normalization across different target profiles as compared to efficiency $\eta$, and its closer relevance to application-level performance as compared to accuracy $\alpha$.

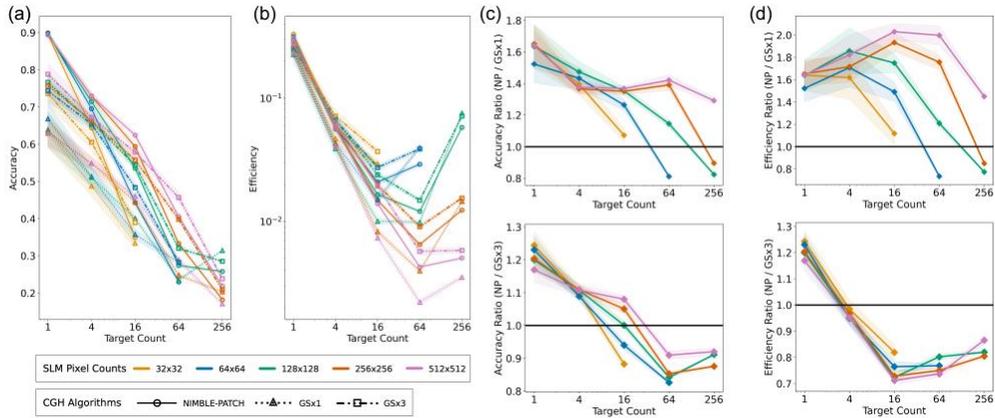

Fig. S2. Sweep results comparing NIMBLE-PATCH to GS for single-frame simulations in terms of the two auxiliary metrics. (a) Accuracy and (b) efficiency as a function of target count $N$ across SLM pixel counts $F \times F$. (c) Accuracy ratios and (d) efficiency ratios between NIMBLE-PATCH and the two GS algorithms (top: GSx1, bottom: GSx3). 95% confidence intervals for the mean value of the curves are shown as shaded regions for each plot.

## 2. Experimental data processing and quantification

For each evaluated phase mask at a given SLM pixel count/total target count $(F, T)$ condition, camera exposure time was adjusted to avoid saturation, and an image z-stack was obtained by averaging two acquisition runs at a z-stage step size corresponding to $1/12^{th}$ of the Abbe axial spot size. In order to correct for aberrations impacting focus tuning as well as misalignment between the stage-mounted camera plane and the optical axis in the acquired z-stacks, a single-point on-axis focus tuning calibration was performed in order to record both the measured depth and optical axis position (in camera pixel coordinates) corresponding to various targeted depths. The recorded deviation between measured and intended depths, which was accurately accounted for with a third-order polynomial fit as shown in Fig. S3, can likely be attributed to target depth-dependent spherical aberrations from the beamsplitter and Fourier-transforming lens.

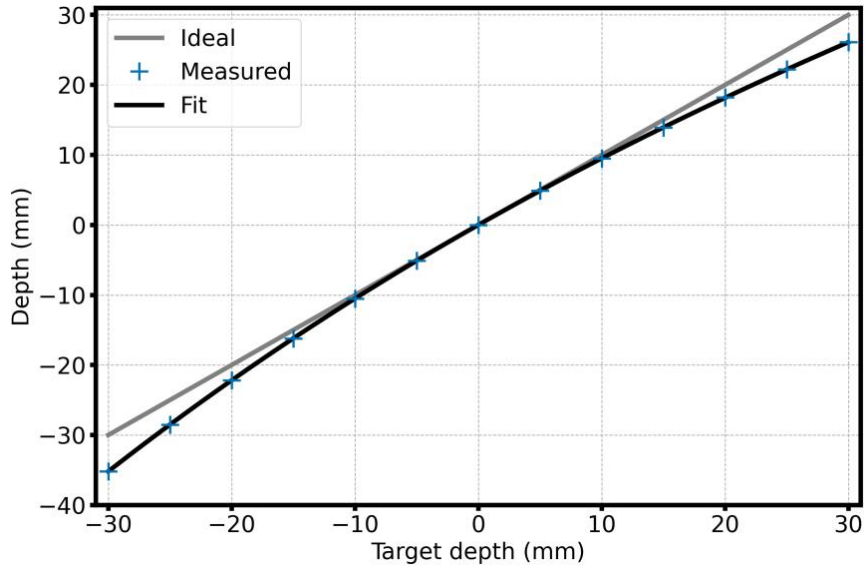

Fig. S3. Intended target depth plotted against recorded peak intensity depth from on-axis focus tuning calibration for experimental acquisitions. Measured deviation is accurately captured by third order polynomial fit.

Using linear regression of the recorded optical axis coordinates against measured depths, the acquired z-stacks were computationally realigned by centering each image slice to the optical axis then cropping to the lateral FoV bounds of the evaluated $(F, T)$ pair. Peak intensity locations along $x'$, $y'$, and $z'$ were subsequently identified, and a linear sum assignment was employed to assign each targeted location to an identified peak using distance as a cost matrix. Full widths at half maximum (FWHMs) along $x'$, $y'$, and $z'$ were also recorded for each peak as part of the peak identification process. A target point was considered non-identified (red ellipses in Fig. 5) if the assigned peak is located more than 3 FHWMs away along each dimension as given by: $FWHM_{x,y} = 1.02 w_{x,y}$ and $FWHM_z = 0.9 w_z$. The percentage of successfully identified targets and resulting position errors between identified targets and their assigned peak locations are shown in Fig. S4(a) and FigS4(b), respectively. In order to compare target optical power across $(F, T)$ conditions, recorded camera pixel values were scaled in accordance with acquisition exposure times, and irradiances were evaluated across target size disk regions set by patch dimensions and defined in $I(x', y', z')$.

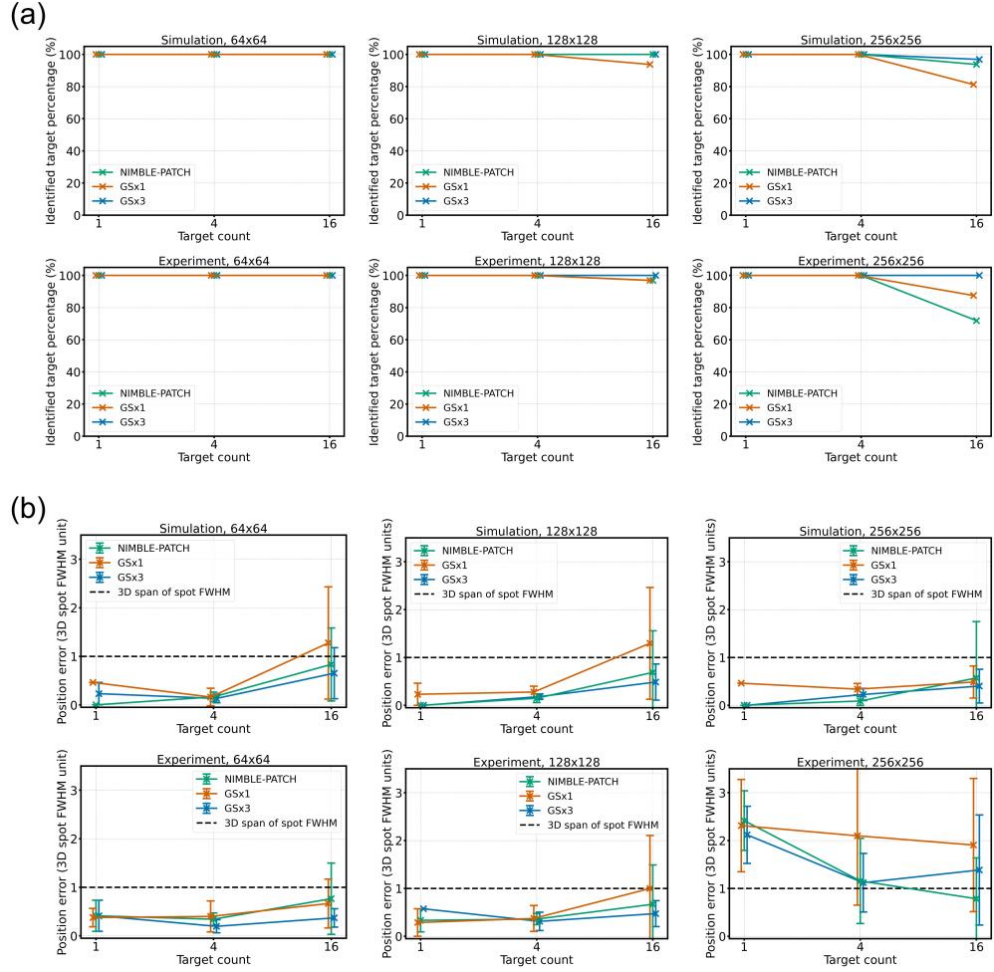

Fig. S4. (a) Percentage of successfully identified targets across both point cloud volumes evaluated at each SLM pixel count/total target count $(F, T)$ condition in simulation and experiment. (b) Position error between identified targets and their matched peak intensity locations across both point cloud volumes evaluated at each $(F, T)$ condition in simulation and experiment. Error bars represent standard deviation and dashed line reference corresponds to the 3D diagonal span of the bounding box formed by theoretical FHWM spot dimensions.